\documentclass[10pt, journal, letterpaper]{IEEEtran} 

\AtBeginDocument{%
  \providecommand\BibTeX{{%
    \normalfont B\kern-0.5em{\scshape i\kern-0.25em b}\kern-0.8em\TeX}}}
    
\usepackage[T1]{fontenc}
\usepackage{graphicx}
\usepackage{float}
\graphicspath{ {figs/} }
\usepackage{amsmath}
\usepackage{dsfont}
 
\usepackage[ruled,vlined]{algorithm2e}
\usepackage[font=small]{caption}
\usepackage{subcaption}
\usepackage[export]{adjustbox}
\usepackage{siunitx}
\usepackage{colortbl}
\usepackage{algorithmic}
\usepackage[thinlines]{easytable}
\usepackage{siunitx}
\usepackage[normalem]{ulem}
\usepackage{amsfonts}
\usepackage{multirow,verbatim}
\usepackage[normalem]{ulem}
\usepackage[dvipsnames]{xcolor}
\usepackage{url}

\newcommand{\name}{TITAN}

\newcolumntype{b}{>{\columncolor{blue!10}}c}
\newcolumntype{y}{>{\columncolor{yellow!10}}c}
\newcolumntype{d}{>{\columncolor{red!7}}c}

\SetKwInput{KwInput}{Input}                
\SetKwInput{KwOutput}{Output}              

\begin{document}


\title{\name: \underline{T}win-\underline{I}nformed \underline{T}opology \underline{A}daptatio\underline{N} for LAWN-enabled D2C Communication}


\author{	\IEEEauthorblockN{
\small
       Talip Tolga Sarı\IEEEauthorrefmark{1}, Rameez Ahmed\IEEEauthorrefmark{2}, Abdullah Al Noman\IEEEauthorrefmark{3}, Gökhan Seçinti\IEEEauthorrefmark{1}\IEEEauthorrefmark{4}, Chris Dick\IEEEauthorrefmark{5} and Debashri Roy\IEEEauthorrefmark{3}
        }\\
\IEEEauthorblockA{
\small 
\IEEEauthorrefmark{1}Istanbul Technical University, Turkey; \IEEEauthorrefmark{2}The Mathworks Inc., USA;
\IEEEauthorrefmark{3}The University of Texas at Arlington, USA; \\
\IEEEauthorrefmark{4}ITU-BTS Digital Twin Application and Research Center, Turkey;
\IEEEauthorrefmark{5}NVIDIA Corporation, USA \\
Emails: \footnotesize{\texttt{sarita@itu.edu.tr, rrasheed@mathworks.com, abdullahal.noman@uta.edu,} \\ \texttt{secinti@itu.edu.tr, cdick@nvidia.com, debashri.roy@uta.edu}}
}
}

\maketitle

\begin{abstract}



Low-Altitude Wireless Networks (LAWN) are transforming the low-altitude airspace into a mission-driven, dynamically reconfigurable 3D network fabric for safety-critical and public-safety operations. In parallel, Direct-to-Cell (D2C) satellite access can rapidly restore connectivity after disasters, yet dense urban blockages make the satellite-to-ground link unreliable for many users. To overcome this, we leverage the LAWN aerial layer and form an adaptive low-altitude relay topology where Unmanned Aerial Vehicles (UAVs) act as D2C-assisted aerial relays for obstructed ground users. We introduce TITAN, a twin-informed topology adaptation framework that builds a high-fidelity Digital Twin (DT) of the affected urban area and performs site-specific, ray-traced air-to-ground channel modeling via Sionna RT. This informs a Bayesian optimization process that adapts the aerial topology to maximize coverage and Quality of Service (QoS) for ground users by using UAVs as optimal D2C relays. Extensive system-level simulations with Sionna show that TITAN consistently outperforms the baselines and delivers +32.2\% user coverage, +64.9\% system sum-rate, and +49.3\% fairness over the state-of-the-art (SOTA) that employ heuristic placement or statistical channel approximations. To support further research in resilient network design, we open-source the codebase of the TITAN framework.



\end{abstract}

\begin{IEEEkeywords}
Low-Altitude Wireless Networks, Digital Twin, D2C, UAV Placement, Topology Adaptation, Bayesian Optimization, Ray Tracing, 6G.
\end{IEEEkeywords}
\vspace*{-5pt}

\section{Introduction}
\label{sec:intro}



\begin{figure}[t!!]
    \centering
        \includegraphics[width=0.9\linewidth]{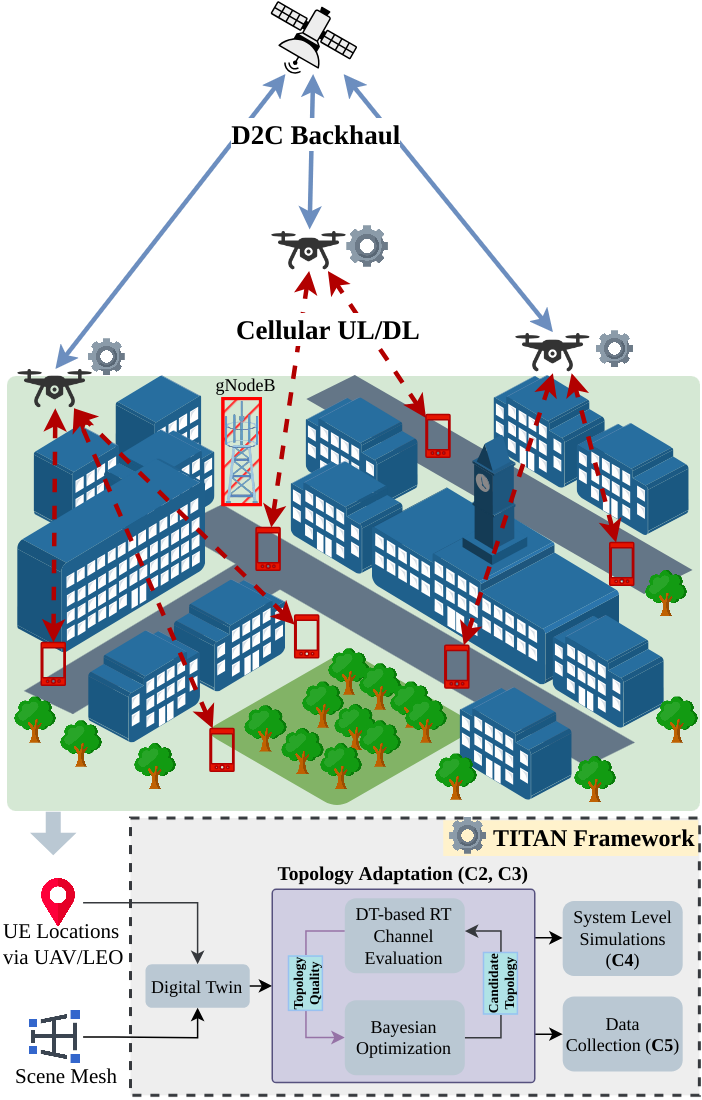}
        \vspace{-5pt}
        \caption{Proposed ray-traced Low-Altitude Wireless Network (LAWN)-enabled Digital Twin (DT) framework highlighting the core contributions: The DT ingests environment mesh and UE locations to compute precise channel impulse responses \textbf{(C1)}, derives optimal 3D UAV placements via Bayesian optimization \textbf{(C2)} to maximize coverage/QoS, and enables reactive topology adaptation \textbf{(C3)} to environmental changes. Finally, it performs system-level simulation \textbf{(C4)} and data collection \textbf{(C5}).}
      \vspace{-20pt}
    \label{fig:intro}
  
\end{figure}

In the immediate aftermath of natural or man-made disasters, the single most critical asset for an effective response is a functioning communication network. The ability to dispatch first responders, issue public safety alerts, assess damage, and connect separated families is entirely dependent on the flow of information \cite{ali2021review}. However, the very infrastructure we rely on is often the first casualty. Disasters can cause widespread power outages, sever fiber optic backhaul lines, and physically destroy cellular base stations. This collapse severely impairs information flow and coordination, severely hampering situational awareness and complicating rescue efforts at a time when coordination is paramount \cite{anjum2023space, ariman2023energy}. The urgent need for rapidly deployable, resilient, and high-performance communication solutions to operate in these contested environments has never been greater, motivating the TITAN framework illustrated in Fig.~\ref{fig:intro}.

\begin{figure}
    \centering
    \includegraphics[width=0.99\linewidth]{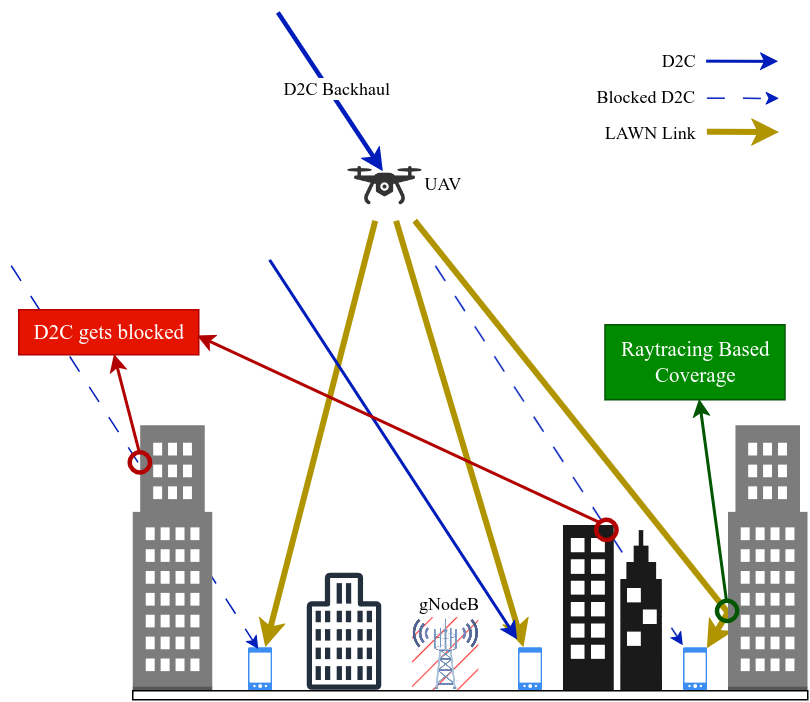}
    \vspace{-10pt}
    \caption{D2C signals directly reaching UEs are frequently blocked or severely attenuated. Instead, UAVs leverage stronger, more reliable D2C satellite links due to superior antenna alignment and guaranteed LoS, serving as aerial relays. }
    \vspace{-20pt}
    \label{fig:d2c}
\end{figure}

\noindent{\bf The Promise and Peril of D2C Connectivity:}
A transformative technology poised to address this challenge is D2C communication, which connects standard user devices directly to Low Earth Orbit (LEO) satellite constellations. This paradigm offers the revolutionary potential for near-instantaneous service restoration over vast areas without any reliance on local ground infrastructure. However, while D2C is effective in open rural areas, its viability in dense urban environments, where a significant portion of the population resides, is severely compromised. The satellite-to-ground link requires a clear Line-of-Sight (LoS) or near-LoS path, \textit{but in urban canyons,} this path is almost always obstructed by buildings and other structures, as illustrated in Fig. \ref{fig:d2c}. The steep elevation angles required to connect with LEO satellites mean that signals must penetrate multiple walls or be perfectly reflected, leading to crippling attenuation, polarization mismatch, and multipath fading that render the D2C link unusable for most ground-level users \cite{chen2020vision}.

\noindent{\bf From Architectural Solution to an Adaptation Problem: The Role of Low-Altitude Wireless Networks (LAWNs) and Digital Twins:}
To bridge this critical gap, an architectural solution is needed: a LAWN. Within a LAWN, the aerial layer, comprised of agile assets like UAVs, can serve as an on-demand relay network~\cite{LAWN_survey}. A UAV  can maintain a strong and stable LoS backhaul link to an overhead satellite and then relay that connection to users on the ground, effectively circumventing the urban blockages that plague D2C~\cite{zhang2018analysis}. This LAWN-enabled D2C model fundamentally reframes the problem. The question is no longer about the static availability of a link but about the dynamic deployment of an adaptive network topology. The challenge becomes one of intelligence and optimization: How can we autonomously place the UAVs to form the most effective relay topology in a complex, unpredictable, and partially destroyed environment?

This intelligent placement is impossible without a deep, site-specific understanding of the radio propagation environment. Conventional network planning tools, which rely on overly-simplified stochastic channel models, are inadequate~\cite{charbonnier2020calibration}. They average out the very deterministic, location-specific blockages and reflections that define the urban channel, leading to grossly inaccurate performance predictions and a suboptimal, unreliable aerial topology. To solve this, we must transition from statistical guesswork to deterministic prediction. This necessitates a high-fidelity Digital Twin (DT), as it enables precise environmental modeling and accurate performance forecasting for any candidate network configuration~\cite{kuruvatti2022empowering}.

\noindent{\bf Key Research Challenges:}
Implementing a DT-informed framework for adaptive topology control requires overcoming several significant research challenges:

\begin{itemize}
    \item \textbf{(Ch1) High-Fidelity and Efficient Channel Modeling:} How can we accurately model the site-specific wireless channels across a large urban area? The modeling approach must capture deterministic effects like reflection and diffraction yet remain computationally efficient enough to be used within a real-time optimization loop.
    \item \textbf{(Ch2) Complex and High-Dimensional Topology Optimization:} How do we efficiently find the optimal number and 3D coordinates for a fleet of UAVs? The search space for this problem is vast, continuous, and high-dimensional. Furthermore, the objective function, overall network performance, is a ``black box" that can only be evaluated through computationally expensive ray-tracing simulations, rendering brute-force or exhaustive search methods infeasible.
    \item \textbf{(Ch3) Adjusting to Environmental Dynamics:} How can the aerial topology reactively adapt to the fluid nature of a disaster scenario? A truly resilient network must be able to reconfigure its topology in response to dynamic changes, such as the failure of remaining ground-level base stations or shifts in user density.
    \item \textbf{(Ch4) Ensuring Robustness to Real-World Imperfections:} How will the optimized topology perform given the inevitable uncertainties of a real-world deployment? A practical framework must demonstrate robustness against imperfections such as noisy or outdated GPS data for ground users and varying levels of fidelity in the DT itself.
\end{itemize}

\noindent{\bf Our Proposed Solution: The \name~Framework.}
To address these multifaceted challenges, this paper introduces \textbf{\name}, a Twin-Informed Topology Adaptation framework for LAWN-enabled D2C communication. At its heart, \name~ leverages a high-fidelity DT constructed from best available 3D environmental data (e.g., city mesh models, rapidly captured post-disaster aerial surveys) and the last known locations of users. By creating this virtual replica, \name~ replaces inaccurate stochastic models with precise, deterministic ray-traced channel simulations powered by Sionna RT~\cite{sionna}, directly addressing \textbf{(Ch1)}. This accurate channel knowledge becomes the input to a powerful Bayesian optimization engine, which intelligently and efficiently explores the high-dimensional search space to find a near-optimal aerial topology without requiring an exhaustive search, thus solving \textbf{(Ch2)}. The framework is designed to be dynamic and adaptive, enabling topology reconfiguration in response to user mobility and evolving link conditions \textbf{(Ch3)}. Finally, we rigorously test the system under imperfect digital twin representations and erroneous situational awareness to validate its real-world robustness mentioned in \textbf{(Ch4)}.

\noindent{\bf Summary of Contributions:}
The primary contributions of this work are as follows:

\begin{itemize}
    \item \textbf{(C1) High-Fidelity LAWN Modeling Framework:} We design and implement a novel DT framework that integrates a state-of-the-art ray-tracing engine and system-level simulator, Sionna RT~\cite{sionna}. This provides an end-to-end tool for accurately modeling and evaluating the performance of adaptive LAWN topologies in complex, realistic urban environments. For realistic LEO dynamics, we incorporate Starlink Two-Line Element (TLE)-based satellite ephemerides into the D2C evaluation pipeline.
    \item \textbf{(C2) Efficient Bayesian-based Topology Optimization:} We formulate the aerial network design as a complex, black-box optimization problem. We then solve it efficiently using Bayesian optimization to jointly determine the optimal number and 3D locations of UAVs, maximizing a composite utility function of network coverage, total capacity, and user fairness.
    \item \textbf{(C3) Adaptation to Environment Dynamics:} We analyze the reactive deployment of the \name~framework across diverse disaster scenarios, capturing various environmental dynamics. Our findings demonstrate that such dynamic placement significantly enhances QoS compared to sole reliance on direct satellite access, which suffers from severe urban blockages.
    \item \textbf{(C4) Comprehensive Robustness Analysis:} We conduct an extensive evaluation of \name's performance under non-ideal, real-world conditions. We demonstrate the framework's resilience to significant uncertainties, including errors in user location data and the trade-offs between DT fidelity and computational cost.

    \item \textbf{(C5) Open-Source Framework for Community Research:} We release our TITAN framework as an open-source tool in \cite{twist}. This contribution enables reproducible research and provides the community with a powerful platform for developing and testing new algorithms for network adaptation, as well as for generating large-scale datasets (e.g., channel impulse responses) to train future data-driven and ML-based solutions. Moreover, to ensure reproducibility and real-world fidelity, the framework incorporates actual TLE sets from operational Starlink satellites.
\end{itemize}

The remainder of this paper is organized as follows. Section~II reviews related work on aerial relay networks, Digital Twins, and D2C satellite communications. Section~III presents the TITAN system architecture, including the network model and problem formulation. Section~IV details the proposed TITAN framework, encompassing the ray-tracing based Digital Twin implementation and the Bayesian optimization approach for aerial topology adaptation. Section~V describes the evaluation methodology, simulation parameters, and performance metrics. Section~VI provides comprehensive experimental analysis and robustness evaluation under various deployment scenarios. Finally, Section~VII concludes the paper with a summary of contributions and directions for future research.



\section{Related Works}

The literature review highlights that although LAWN architectures, DT-driven network optimization, and UAV placement methods are well-explored individually, there remains a significant gap in effectively combining these technologies to address resilient D2C communication specifically in post-disaster urban environments.

\subsection{LAWNs for Public Safety and Disaster Response}

Unlike traditional Space-Air-Ground Integrated Networks (SAGINs) that span from satellites to ground infrastructure~\cite{shang2021computing, Marinho24}, LAWNs focus on the low-altitude aerial layer (typically below 1,000 meters), where agile assets such as UAVs can be rapidly deployed to establish on-demand connectivity \cite{yuan2025ground}. This layer is attractive for disaster response because it can be positioned and repositioned quickly, often preserving stronger LoS links to ground users across complex urban terrain \cite{jin2026advancing}. In these architectures, UAVs typically operate as temporary aerial relays or base stations that restore coverage when terrestrial infrastructure is degraded.

A core challenge is that effective UAV placement is a high-dimensional, continuous 3D problem that becomes harder as the fleet grows. Many classical approaches rely on analytical or simplified stochastic channel models (e.g., 3GPP TR38.901 urban macrocell), which approximate LoS availability using statistical averages~\cite{bor2016efficient, 9016553, 9621129}. In dense urban canyons, such averaging can miss site-specific blockage and reflection patterns that dominate performance, which limits reliability. Recent learning-based placement methods can improve search efficiency, but they may still be too evaluation-heavy for real-time response or remain constrained by imperfect channel representations in fast-changing disaster conditions~\cite{jin2026advancing}.

While optimal placement is critical, on-board energy constraints remain a practical bottleneck for sustained public-safety operations using UAVs. However, the state-of-the-art is rapidly evolving to mitigate this inherent constraint through innovations in in-flight energy replenishment and autonomous logistics. Emerging solutions include inter-UAV wireless charging and automated hot-swapping mechanisms. Notably, magnetic resonance coupling-based Wireless Power Transfer (WPT) has demonstrated transmission efficiencies exceeding 90\% at short distances, enabling feasible autonomous charging-station integration~\cite{pang2023lightweight}. Furthermore, automated battery management and coordinated UAV swapping support persistent coverage by seamlessly replacing depleted units with charged counterparts~\cite{ure2014automated, burdakov2017optimal}. Given these advancements in hardware and energy management, our work differentiates itself by focusing on the algorithmic challenge of site-specific, DT-informed placement: we couple deterministic (ray-traced) channel evaluation with efficient black-box optimization to adapt the aerial topology to the real urban environment rather than to statistical expectations.

\subsection{DTs for Wireless Network Optimization}
DTs have become central for network optimization, including spectrum sharing, network management, and predictive resource allocation, with integration of ray-traced channels proving valuable for generating realistic training data \cite{alkhateeb2022deepmimo,RAZA2025108144}. Recent DT research emphasizes accurate urban modeling and proactive 6G network planning \cite{10198575,10283539,10999688}. Yet, despite these advancements, existing studies primarily focus on terrestrial networks and seldom apply DT-based optimization for dynamically adapting aerial topologies in disaster-driven D2C contexts, marking a clear gap addressed by our approach.

The application of DTs has extended to aerial networks, where they enable UAV placement optimization through high-fidelity ray-tracing simulations \cite{multidt2025}. However, existing approaches typically focus on either link-level channel estimation \cite{dtddpg2024} or fixed aerial infrastructure \cite{dtdrl2026}, without considering end-to-end system performance. In disaster management scenarios, DTs facilitate rapid infrastructure restoration through real-time network adaptation \cite{dtpower2024}. Furthermore, the importance of cross-layer modeling (spanning PHY, MAC, and network layers) has been recognized for accurate wireless DTs \cite{gblm2024}.

\subsection{Limitations in State-of-the-Art and Our Contributions}
While current literature thoroughly explores each individual component, critical limitations persist when directly applied to resilient D2C communication scenarios outlined earlier. Most notably, existing works employ oversimplified stochastic channel models rather than the required deterministic urban propagation environments (related to \textbf{Ch1}). Additionally, channel modeling and topology optimization frequently remain decoupled, reducing solution efficacy (related to \textbf{Ch2}). Lastly, current solutions largely neglect dynamic, real-time topology adaptation to rapidly evolving disaster scenarios (related to \textbf{Ch3}). TITAN explicitly targets these challenges by integrating precise ray-traced DT modeling, tightly coupled Bayesian optimization, and adaptive capabilities, thereby significantly enhancing practical effectiveness compared to current state-of-the-art (SOTA) approach~\cite{multidt2025}.

\section{TITAN System Architecture}

We now present the end-to-end system architecture, comprising three functional layers: (i) the physics-based channel modeling layer utilizing ray-traced DTs, (ii) the Bayesian optimization engine for high-dimensional topology search, and (iii) the reactive control mechanism for environmental dynamics. The following subsections detail the algorithmic implementations and mathematical formulations of each component.

\subsection{DT and Ray-traced Channel Model (C1, C5)}
The foundation of \name~ is its ability to accurately predict wireless channel behavior. {Rather than relying on statistical models, we leverage Sionna RT's ray-tracing engine to compute site-specific channel responses within a high-fidelity DT of the disaster environment.}

\subsubsection{DT Construction}
The DT is a virtual replica of the disaster zone, created by fusing two primary data sources:
\begin{enumerate}
    \item \textbf{Environmental Data:} A high-resolution 3D mesh of the urban environment, including buildings, terrain, and foliage. This can be sourced from existing city models or captured post-disaster via aerial surveys.
    \item \textbf{User Equipment (UE) Locations:} The coordinates of ground users, determined via onboard GPS receivers and subsequently broadcasted by the devices to the network, or alternatively retrieved from existing network logs.
\end{enumerate}
This DT serves as the input to a ray-tracing engine, which models the propagation of electromagnetic waves through this specific, complex environment.

\subsubsection{Ray-traced Channel Formulation}
For any given aerial topology, defined by the 3D coordinates of the UAV fleet, the ray tracer computes the precise Channel Impulse Response (CIR) for the link between each UE and each UAV. Assuming channel reciprocity, the computed CIR applies to both uplink and downlink transmissions. The coherent, time-varying CIR for the link between UE $u$ and UAV $b$ is expressed as a sum of discrete multipath components:
\begin{equation}
h_{u,b}(t, \tau) = \sum_{l=1}^{L_{u,b}} \alpha_{l,u,b}(\tau) \delta(t - \tau_{l,u,b}(\tau)),
\end{equation}
where $L_{u,b}$ is the total number of distinct propagation paths (LoS, reflected, diffracted), $\alpha_{l,u,b}$ is the complex gain of the $l$-th path, and $\tau_{l,u,b}$ is its corresponding delay. The received power is then calculated as the coherent sum across all paths: $P_{u,b} = |\sum_{l}^{L_{u,b}} \alpha_{l,u,b}|^2$. This allows us to compute the Signal-to-Interference-plus-Noise Ratio (SINR) for UE $u$ associated with its serving UAV $b^*$, considering interference from all other UAVs $b' \neq b^*$:
\begin{equation}
\gamma_{u,b^*} = \frac{P_{u,b^*}}{\sum_{b' \neq b^*} P_{u,b'} + N_0},
\end{equation}
where $N_0$ is the thermal noise power. A UE is considered to have coverage if its SINR exceeds a predefined threshold, $\gamma_{th}$. This high-fidelity, site-specific channel knowledge is the key enabler for intelligent topology adaptation.

\subsection{Aerial Topology Optimization (C2)}
With an accurate method for evaluating network performance, we formulate the core task of \name~ as a black-box optimization problem: finding the best aerial topology to serve the ground users.

\subsubsection{Problem Formulation}
Let $\mathcal{U}=\{1,...,U\}$ be the set of ground UEs and $\mathcal{B}=\{1,...,B\}$ be the set of UAVs forming the aerial layer. The complete topology is defined by the configuration vector $\mathbf{X}=[\mathbf{x}_{1},...,\mathbf{x}_{B}]$, where $\mathbf{x}_{b}=(x_{b},y_{b},z_{b})$ is the 3D coordinate of UAV $b$. Denoting by $\mathcal{U}_{cov}(\mathbf{X}) \subseteq \mathcal{U}$ the set of covered UEs under topology $\mathbf{X}$, a UE $u$ belongs to this set if its SINR ($\gamma_{u,b}(\mathbf{X})$) surpasses a given threshold:
\begin{equation}
u \in \mathcal{U}_{cov}(\mathbf{X}) \iff \gamma_{u,b}(\mathbf{X}) \geq \gamma_{th}.
\end{equation}
The instantaneous channel capacity for UE $u$ on subcarrier $k$ is computed using Shannon's equation:
\begin{equation}
C_{u,k}(\mathbf{X}) = B_{sc} \log_{2}(1 + \gamma_{u,k}(\mathbf{X})),
\end{equation}
where $B_{sc}$ is the subcarrier bandwidth. The total achievable rate (sum capacity) is the sum over all covered UEs and subcarriers:
\begin{equation}
C_{sum}(\mathbf{X}) = \sum_{u \in \mathcal{U}_{cov}(\mathbf{X})} \sum_{k} C_{u,k}(\mathbf{X}).
\end{equation}
The goal is to find the optimal topology $\mathbf{X}^{*}$ that maximizes a composite utility function, $f(\mathbf{X})$:
\begin{equation}
\mathbf{X}^{*} = \arg\max_{\mathbf{X}} f(\mathbf{X}) = C_{sum}(\mathbf{X})^{\omega_{1}} \cdot \text{Cov}(\mathbf{X})^{\omega_{2}} \cdot J(\mathbf{X})^{\omega_{3}}
\label{eq:objective}
\end{equation}
where,
\begin{itemize}
    \item $\text{Cov}(\mathbf{X}) = \frac{|\mathcal{U}_{cov}(\mathbf{X})|}{|\mathcal{U}|}$ is the coverage ratio.
    \item $J(\mathbf{X}) = \frac{(\sum_{b} n_{b}(\mathbf{X}))^2}{|\mathcal{B}| \sum_{b} n_{b}(\mathbf{X})^2}$ is Jain's Fairness Index, with $n_b(\mathbf{X})$ being the number of UEs served by UAV $b$, and $|\mathcal{B}|$ is the number of deployed UAVs.
    \item The weights $\omega_i$ are used to prioritize different KPIs. In this work we choose all of them as $1$.

\end{itemize}
This process also jointly determines the optimal number of UAVs, $|\mathcal{B}|$, by iteratively adding UAVs until a target coverage ratio is met.

\subsubsection{Bayesian Optimization}
Evaluating the objective function in Eq. \ref{eq:objective} is computationally expensive, as it requires running a full ray-tracing simulation for each candidate topology $\mathbf{X}$. An exhaustive search of the high-dimensional, continuous search space is therefore intractable. To solve this efficiently, \name~ employs Bayesian Optimization with a Tree-structured Parzen Estimator (TPE)~\cite{optuna}. TPE is exceptionally well-suited for black-box optimization. It works by building a probabilistic surrogate model of the objective function, mapping topologies to their expected performance. It then uses this model to intelligently select the next topology to evaluate, balancing exploration (testing uncertain but potentially promising regions) and exploitation (refining the best-known solutions). This allows \name~ to converge to a near-optimal and robust aerial topology with a minimal number of computationally expensive simulations.

\subsection{Dynamic Topology Adaptation (C3)}
Disaster scenarios are inherently dynamic. A key feature of \name~is its ability to adapt the aerial topology reactively. When the state of the ground environment changes—for example, if a remaining terrestrial gNodeB fails due to a power outage, the system detects this change. The DT is immediately updated with the new network state (e.g., removing the failed gNodeB from the interference calculation). This triggers a re-optimization process, where \name~re-runs the Bayesian optimization loop to compute a new optimal topology, $\mathbf{X}^*_{new}$, that is tailored to the new conditions. This reactive adaptation ensures the aerial network remains effective as the disaster situation on the ground evolves.

\begin{figure*}
    \centering
    \includegraphics[width=\linewidth]{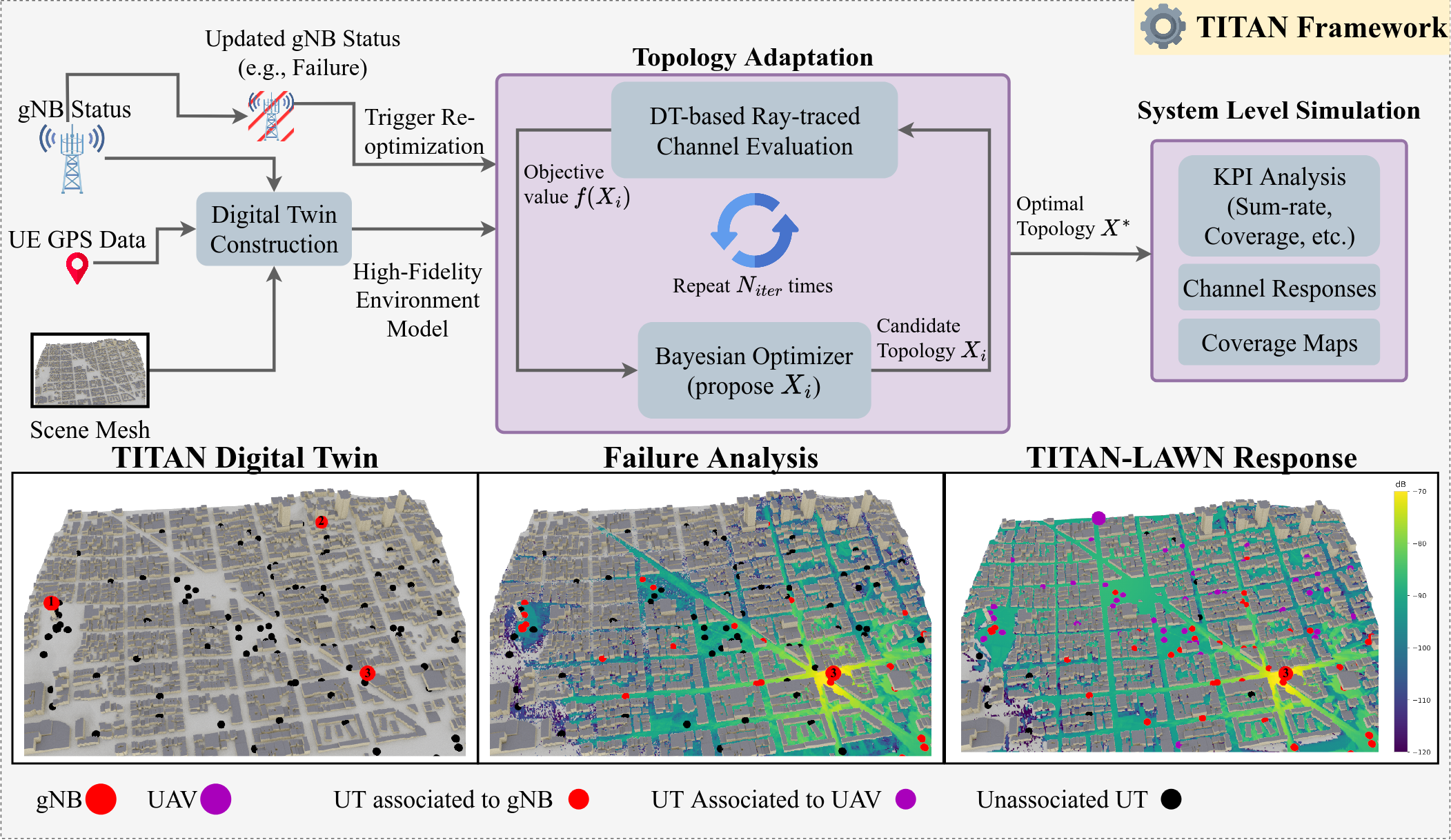}
    \caption{\textbf{The TITAN Framework Workflow and Simulation Example.} The San Francisco mesh includes detailed building models and ground elevation variations to capture realistic urban topography. UEs are positioned 1.5 meters above the local ground surface to represent typical handheld device usage, buildings are modeled as blocks with marble and metal (roof) material types. After obtaining the UE GPS data and gNB information, TITAN creates Digital Twin of the environment. After gNB numbers 1 and 2 fail, TITAN analyzes the environment status and deploys LAWN response. Color bar represents the maximum path gain.}
    \label{fig:framework}
\end{figure*}

\section{The Proposed TITAN Framework}
{
The proposed \name~framework follows a closed-loop, iterative process that leverages the DT to find and deploy a near-optimal aerial topology. This section details the end-to-end workflow of the \name~framework, from initial data ingestion to final performance evaluation and dynamic adaptation. The overall process is illustrated in Fig.~\ref{fig:framework}.

\subsection{Step 1: DT Initialization and Scene Ingestion}
The process begins with the creation of the DT. The \name~framework ingests the necessary environmental and situational data to construct a high-fidelity virtual replica of the disaster zone. This includes:
\begin{itemize}
    \item \textbf{3D Environmental Mesh:} A detailed 3D model of the urban environment is used for DT generation.
    \item \textbf{Initial User and Network State:} The last known locations of ground UEs are populated within the DT. The status of any remaining terrestrial gNodeBs is also registered.
\end{itemize}
This initial state forms the baseline for the topology optimization, priming the DT to evaluate any candidate aerial topology within this specific environment. This step directly addresses our goal of high-fidelity, site-specific modeling \textbf{(C1)}.

\subsection{Step 2: Iterative Topology Optimization via Bayesian Engine}
This step forms the core optimization loop of the \name~framework, designed to efficiently solve the complex placement problem \textbf{(C2)}. The process is orchestrated by the Bayesian optimization engine and is detailed in Algorithm 1.
\begin{enumerate}
    \item \textbf{Propose Candidate Topology:} The Bayesian optimizer proposes an initial candidate aerial topology, $\mathbf{X}_i$, which specifies the number of UAVs, $|\mathcal{B}|$, and the 3D coordinate of each, $\mathbf{x}_{b}$.
    \item \textbf{Evaluate Performance in DT:} The proposed topology, $\mathbf{X}_i$, is passed to the DT for a ray-tracing simulation to compute the channel between the proposed UAVs and all ground UEs.
    \item \textbf{Calculate Objective Function:} Using the ray-traced channels, a system-level simulation is run to determine the network KPIs (sum-rate, coverage, fairness), which are used to calculate the value of the objective function, $f(\mathbf{X}_{i})$.
    \item \textbf{Update Optimizer's Model:} The pair $(\mathbf{X}_{i}, f(\mathbf{X}_{i}))$ is fed back to the Bayesian optimizer, which updates its internal probabilistic surrogate model.
    \item \textbf{Iterate:} The optimizer uses its updated model to propose the next, more promising candidate topology, $\mathbf{X}_{i+1}$. The loop repeats for a predefined number of iterations ($N_{iter})$.
\end{enumerate}
Once the loop terminates, the topology with the highest observed objective function value, $\mathbf{X}^{*}$, is selected as the optimal configuration.

\subsection{Step 3: Dynamic Adaptation to Disaster Scenarios}
\name~is designed not as a static planner, but as a dynamic adaptation engine \textbf{(C3)}. The framework can be re-triggered whenever the state of the environment changes. For instance, if a terrestrial base station fails, that information is fed back into the DT. The interference landscape changes, and the optimization loop (Step 2) can be re-run to find a new optimal topology that accounts for the new conditions. This allows \name~to reactively adapt to the evolving nature of a disaster zone.

\subsection{Step 4: Data Logging and Analysis}
Throughout its operation, the \name~framework logs a comprehensive set of data, fulfilling our contribution of providing a rich, open-source platform for further research \textbf{(C5)}. For every evaluated topology, the framework can store:
\begin{itemize}
    \item The full, coherent CIR for every UT-UAV link.
    \item System-level logs, including resource block allocations.
\end{itemize}
This data is invaluable for offline analysis, validating new communication algorithms, and training future machine learning models.

\begin{algorithm}
\caption{\name~Topology Optimization Algorithm}
\begin{algorithmic}[1]
\STATE \textbf{Input:} Environmental Mesh $\mathcal{M}$, UE Locations $\mathcal{P}_U$, Target Coverage $\rho_{target}$, Max Iterations $N_{iter}$
\STATE \textbf{Output:} Optimal Topology $\mathbf{X}^{*}$
\STATE Initialize DT with $\mathcal{M}$ and $\mathcal{P}_U$
\STATE Initialize UAV count $|\mathcal{B}| \leftarrow 1$
\STATE Initialize Bayesian Optimizer $\mathcal{O}$
\STATE Initialize best observed performance $f_{best} \leftarrow -\infty$
\STATE Initialize optimal topology $\mathbf{X}^* \leftarrow \emptyset$ 
\WHILE{Coverage($\mathbf{X}^*$) $< \rho_{target}$}
    \FOR{$i=1$ to $N_{iter}$}
        \STATE Propose candidate topology $\mathbf{X}_i$ from $\mathcal{O}$ for $|\mathcal{B}|$ UAVs
        \STATE Evaluate $f(\mathbf{X}_i)$ using DT ray-tracing (Eq. 6) 
        \STATE Update optimizer: $\mathcal{O} \leftarrow \mathcal{O} \cup \{(\mathbf{X}_i, f(\mathbf{X}_i))\}$ 
        \IF{$f(\mathbf{X}_i) > f_{best}$} 
            \STATE $f_{best} \leftarrow f(\mathbf{X}_i)$ 
            \STATE $\mathbf{X}^* \leftarrow \mathbf{X}_i$ 
        \ENDIF
    \ENDFOR
    \IF{Coverage($\mathbf{X}^{*}$) $<\rho_{target}$} 
        \STATE $|\mathcal{B}|\leftarrow|\mathcal{B}|+1$ \COMMENT{Increment UAV count} 
        \STATE Reset optimizer $\mathcal{O}$ for new search dimension 
    \ENDIF
\ENDWHILE
\STATE \textbf{return} $\mathbf{X}^{*}$ 
\end{algorithmic}
\end{algorithm}

}
\section{Evaluation Methodology}
{
This section details the simulation environment, the baseline methods used for comparison, the key performance metrics, and the specific experimental scenarios designed to rigorously test each of \name's core contributions.

\subsection{Simulation Environment and Parameters}
Our experiments are conducted using a simulation platform built on open source \textbf{Sionna (v1.2)}, a library for link and system level simulations based on TensorFlow and the Mitsuba renderer, for our analysis~\cite{sionna}. The core of our environment is the Sionna RT ray-tracing engine, which enables physics-based channel modeling~\cite{sionna}. 

\subsubsection{Scenario Setup}
The physical environment is modeled using a detailed 3D mesh of a section of San Francisco, as shown in Fig.~\ref{fig:framework}. This mesh provides realistic urban topography, including building structures and ground elevation changes. We randomly distribute 100 UEs across the outdoor areas of the map, with their initial positions assumed to be known via GPS, as this simulated population explicitly represents a deployment of first responders and Emergency Medical Services (EMS) personnel operating within the disaster-stricken urban sector. In scenarios involving terrestrial infrastructure, we place three gNBs at fixed locations to serve as the baseline cellular network. To determine the available LEO satellite resources for our evaluation, we analyzed Starlink satellite visibility over San Francisco using TLE orbital data. We evaluated three elevation angle thresholds (60, 70, and 80 degrees) to identify satellites with favorable geometric conditions for urban coverage. As shown in Fig.~\ref{fig:satellite_analysis}, the number of visible satellites varies over a 24-hour period, with an average of 6 satellites exceeding the 60 degrees at a given time.

\begin{figure}[hbt]
    \centering
    \begin{subfigure}[b]{0.24\textwidth}
        \centering
        \includegraphics[width=\textwidth]{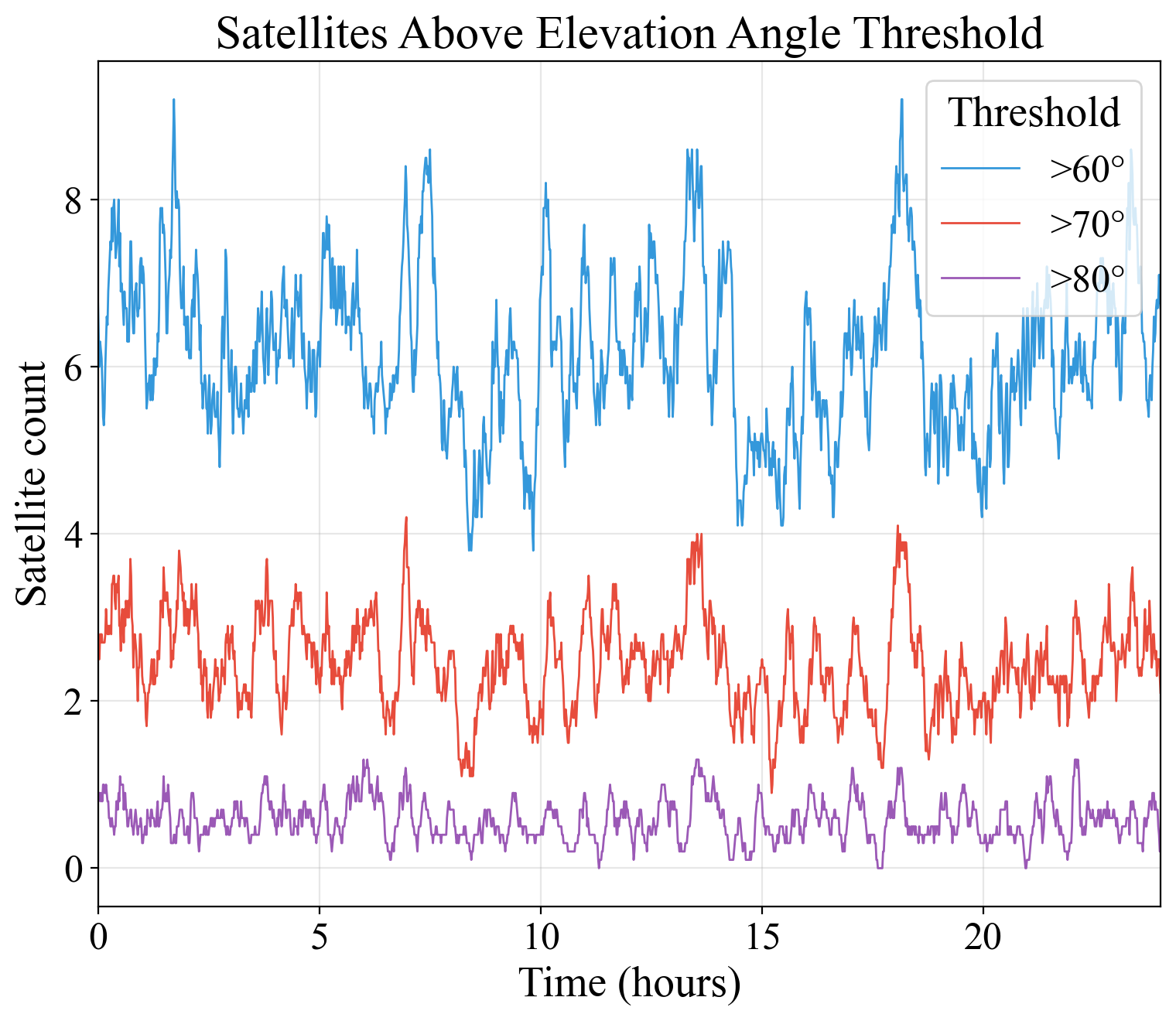}
        \caption{Elevation angle analysis.}
        \label{fig:satellite_counts}
    \end{subfigure}
    \begin{subfigure}[b]{0.24\textwidth}
        \centering
        \vspace{-10pt}\includegraphics[width=\textwidth]{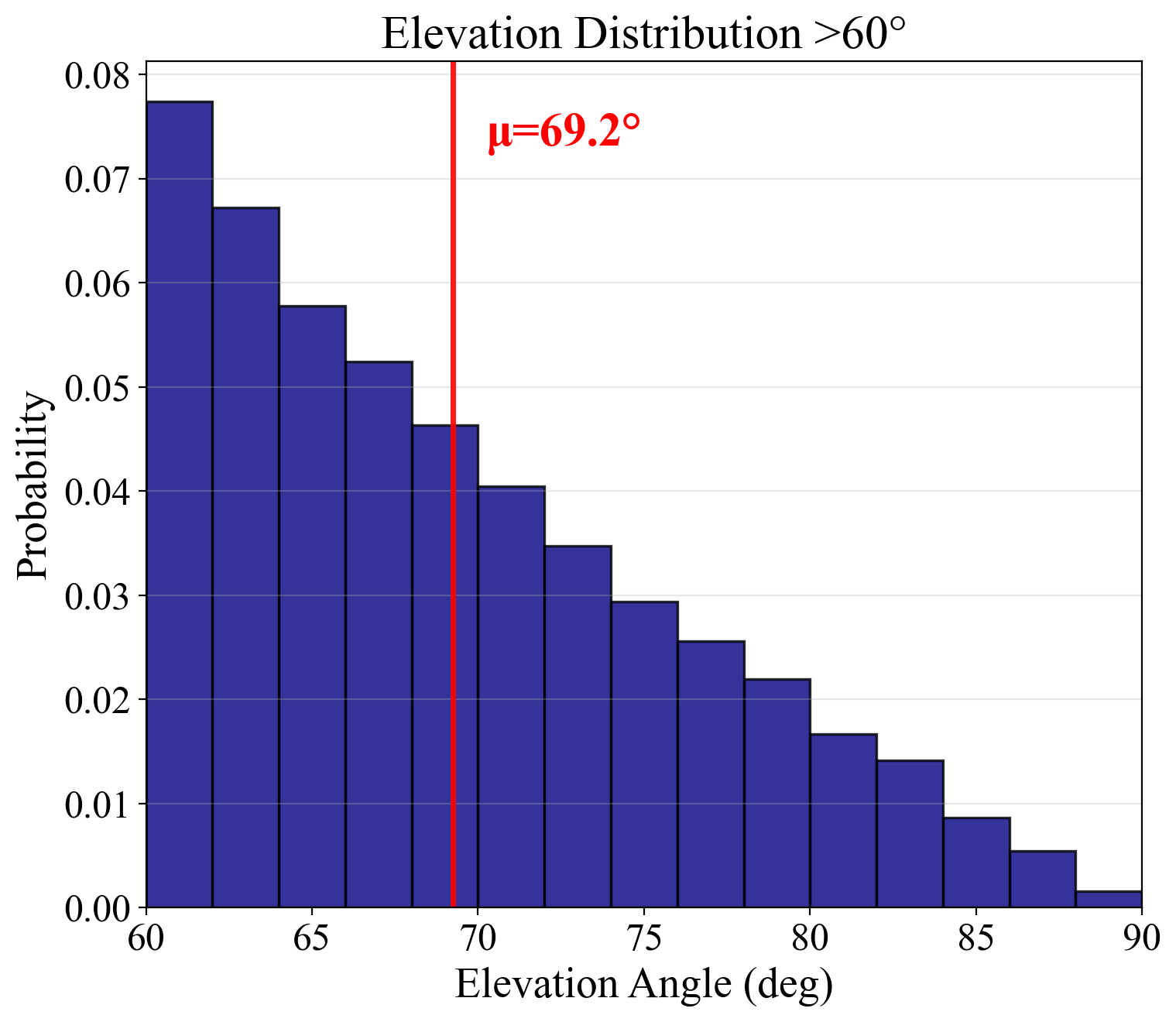}
        \caption{Elevation angle distribution.}
        \label{fig:satellite_pdf}
    \end{subfigure}
    \caption{Starlink satellite visibility analysis over San Francisco for 24-hour window using TLE data with elevation thresholds.}
    \label{fig:satellite_analysis}
\end{figure}

\subsubsection{System-Level Simulation}
For each evaluated topology, we run full system level simulations with Sionna~\cite{sionna}. The simulation incorporates a complete protocol stack, including:
\begin{itemize}
    \item \textbf{Scheduling:} A Proportional-Fair (PF) scheduler is used to allocate resources among users, balancing network throughput and user fairness. We modify the Sionna's implementation to assign resource blocks sized at 12 subcarriers and ensure that the number of unassigned UEs in a slot does not exceed the number of available antennas.
    \item \textbf{Precoding and Equalization:} We employ Regularized Zero-Forcing (RZF) precoding at the transmitter (UAVs and gNBs) and Linear Minimum Mean Square Error (LMMSE) equalization at the UEs to manage multi-user interference.
    \item \textbf{Link Adaptation:} Adaptive Modulation and Coding (AMC) is used based on HARQ feedback to select the appropriate transport block size for the given channel conditions, targeting a block error rate (BLER) of 10\%.
\end{itemize}
The key parameters for our simulation are summarized in Table~\ref{tab:sim_parameters}.

\begin{table}[ht]
\centering
\caption{Simulation Parameters}
\begin{tabular}{ll}
\hline
\textbf{Parameter} & \textbf{Value} \\
\hline
Carrier Frequency & $2$ GHz \\
Subcarrier Spacing & $30$ kHz \\
Bandwidth & $20$ MHz ($5$ MHz for D2C) \\
Total optimization steps ($N_{iter}$) & 150 \\
UAV altitude limits & 50m-1000m \\
Transmit Power (BS, UAV) & $23$ dBm \\
Noise Temperature & $290$ K \\
EIRP (D2C) & $85$ dBm \\
Number of Satellites & 6 \\
Satellite Distance & 600 km \\
gNodeB Antenna Array & $4 \times 4$ Planar Array \\
UE Antenna Array & Single-element isotropic \\
Ray-Tracing Max Depth & $8$\\
Samples per Antenna & $10^6$ \\
Scheduler & Proportional Fair (PF) \\
Precoding & Regularized Zero-Forcing (RZF) \\
Post-Equalization & LMMSE \\
\hline
\end{tabular}
\label{tab:sim_parameters}
\end{table}

\subsection{Competing Methods for Comparison}
To rigorously evaluate \name's performance, we compare it against three distinct baseline approaches that represent alternative strategies for providing coverage in a disaster scenario.

\noindent\textbf{(a) Traditional D2C:} This baseline represents a non-LAWN approach where UEs attempt to connect directly to the LEO satellites. Due to the extreme distance, ray tracing is not feasible. We model this with a simple LoS-only link, where a connection is only possible if an unobstructed path exists between the UE and the satellite at a given elevation angle. This serves as a benchmark to quantify the severity of urban blockage.

\noindent\textbf{(b) Baseline-1:} This baseline provides a lower bound for any aerial-relay solution. It uses the same number of UAVs selected by \name, but places them uniformly at random in 3D within the allowed deployment volume above the target area (same horizontal boundaries and altitude limits). This isolates the benefit of \emph{intelligent} placement over merely having UAVs in the air.

\noindent\textbf{(c) Baseline-2:} This baseline represents optimization \emph{without} a high-fidelity Digital Twin. It keeps the same Bayesian optimization engine as \name, but replaces ray-traced, site-specific channel evaluation with the stochastic 3GPP TR38.901 Urban Macrocell (UMa) channel model. This comparison isolates the value of deterministic, environment-aware DT-driven channel modeling.

\noindent\textbf{(d) Baseline-3:} Here, we use the same Bayesian optimization and ray-traced DT channel evaluation as \name, but the objective is limited to maximizing the sum-capacity across all users. Unlike \name, which uses a composite utility function that balances coverage and fairness, this baseline reflects a pure throughput-maximization approach. This allows us to isolate the specific effect of our utility design on the final UAV placements and resulting performance.

\noindent\textbf{(e) State-Of-The-Art (SOTA)~\cite{multidt2025}:} We implement the UAV placement method of~\cite{multidt2025} using their recommended settings, but with two changes: the area-of-interest (AoI) radius is set to $r_{k}=800\,\mathrm{m}$ and the minimum separation distance to $d_{\min}=500\,\mathrm{m}$, as tuned to match our simulation area. Only the placement algorithm is evaluated, not their power allocation, since our setup already uses MIMO and each UAV serves users on orthogonal resources, as there is no inter-UAV interference. Also, instead of multiple AoIs, we treat the entire region as a single area-of-interest, with users randomly sampled throughout the area for fair comparison.


\noindent\textbf{(f) \name~(Proposed Method):} Our proposed approach employs precise ray-tracing simulations for accurate and deterministic SINR estimation, enabling robust UAV placement optimization. 


\subsection{Performance Metrics}
We evaluate the performance of each method using three principal metrics that align with our optimization objectives:
\begin{enumerate}
    \item \textbf{Sum-Rate (Mbps):} The aggregate data throughput achieved by all successfully served users in the network. This metric reflects the overall network capacity and service quality.
    \item \textbf{Coverage Ratio:} The percentage of total UEs that can successfully establish and maintain a connection with the network (i.e., achieve an SINR above the target threshold $\gamma_{th}=5$ dB).
    \item \textbf{Jain's Fairness Index:} A measure of how equitably network resources are distributed among the served users, calculated based on their individual data rates. This ensures that high sum-rate is not achieved by serving only a few users with excellent channel conditions.
\end{enumerate}

\subsection{Evaluation Scenarios}
We have designed a set of targeted experiments to validate each of the core contributions claimed in Section~\ref{sec:intro}.

\subsubsection{Scenario 1: Baseline Performance in Full Infrastructure Collapse}
This scenario evaluates the core effectiveness of \name's topology optimization \textbf{(C2)} and the benefit of its high-fidelity modeling \textbf{(C1)}. We simulate a total collapse of the terrestrial network (all gNBs are disabled) and task each competing method with providing coverage. By comparing \name~against the Random and Bayesian (TR38.901) baselines, we can quantify the gains from both intelligent placement and accurate channel modeling.

\subsubsection{Scenario 2: Dynamic Adaptation to Partial Failures}
This scenario is designed to validate TITAN's ability to adapt to a dynamic disaster environment \textbf{(C3)}. The simulation starts with all three terrestrial gNBs active. We then progressively disable the gNBs one by one, simulating cascading infrastructure failures. After each failure, TITAN is triggered to re-optimize the aerial topology. We measure the network's ability to maintain service continuity and recover performance through this reactive adaptation.

\subsubsection{Scenario 3: Robustness to Imperfect DT}
This scenario evaluates the framework's robustness to real-world imperfections \textbf{(C4)}. The evaluation is split into two parts:
\begin{itemize}

    \item \textit{Impact of DT Fidelity:} We investigate the trade-off between model accuracy and computational cost. \name's optimization is run using a low-fidelity DT (1\% of the initial amount of rays), but the final, resulting topology is evaluated in a high-fidelity simulation. This measures the performance degradation when operating with a computationally cheaper DT.
    \item \textit{Impact of UE Location Error:} We simulate the effect of inaccurate GPS data by adding Gaussian noise with varying standard deviations (e.g., 1m, 5m, 10m) to the UE locations used during the optimization phase. The final topology is then evaluated against the true, error-free UE locations to measure performance degradation.
\end{itemize}

}

\begin{figure*}[hbt]
    \centering
    \begin{subfigure}[b]{0.45\textwidth}
        \centering
        \includegraphics[width=\textwidth]{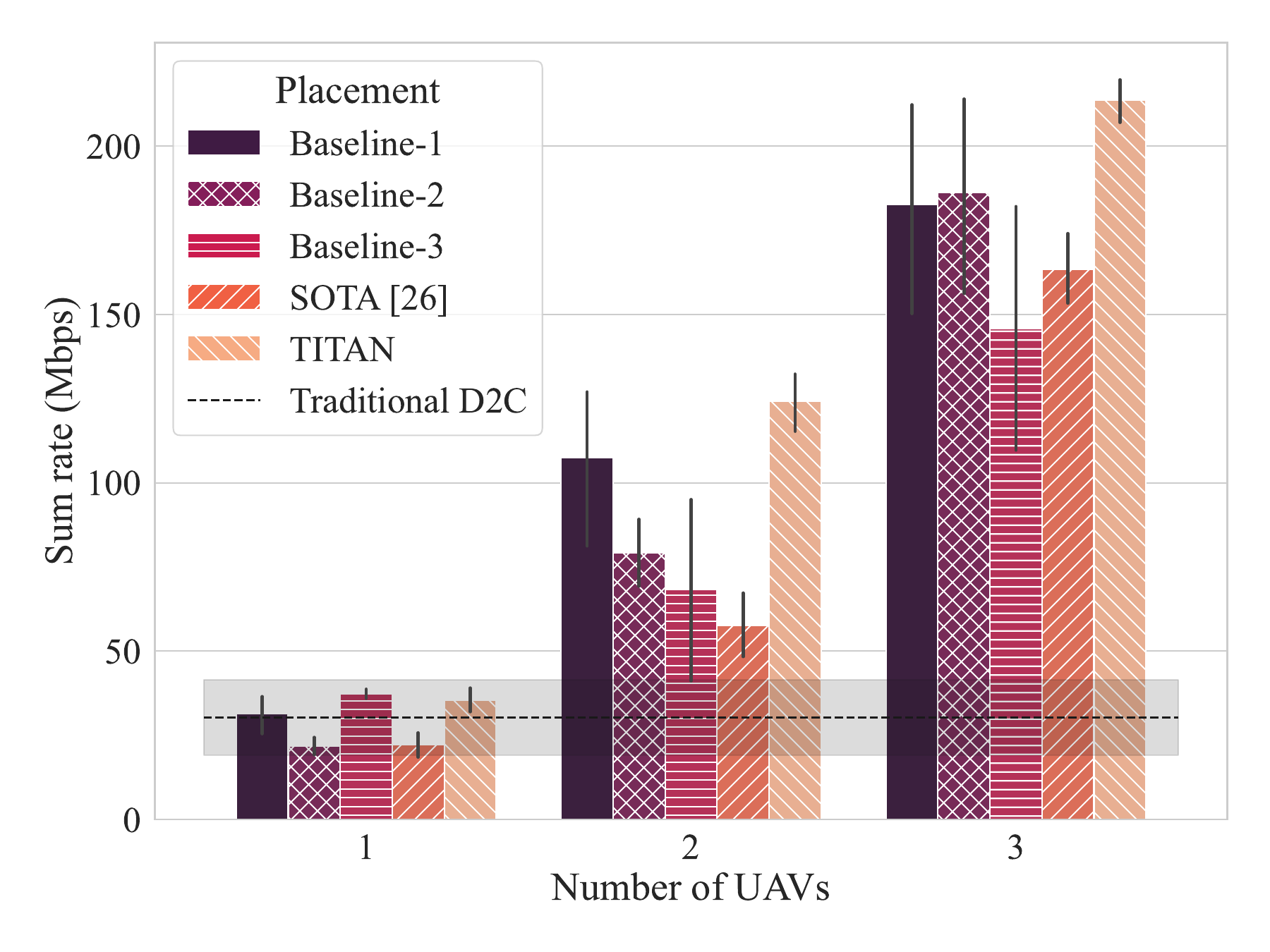}
        \caption{Sum-rate comparison.}
        \label{fig:qos_sumrate}
    \end{subfigure}
    \begin{subfigure}[b]{0.45\textwidth}
        \centering
        \vspace{-10pt}\includegraphics[width=\textwidth]{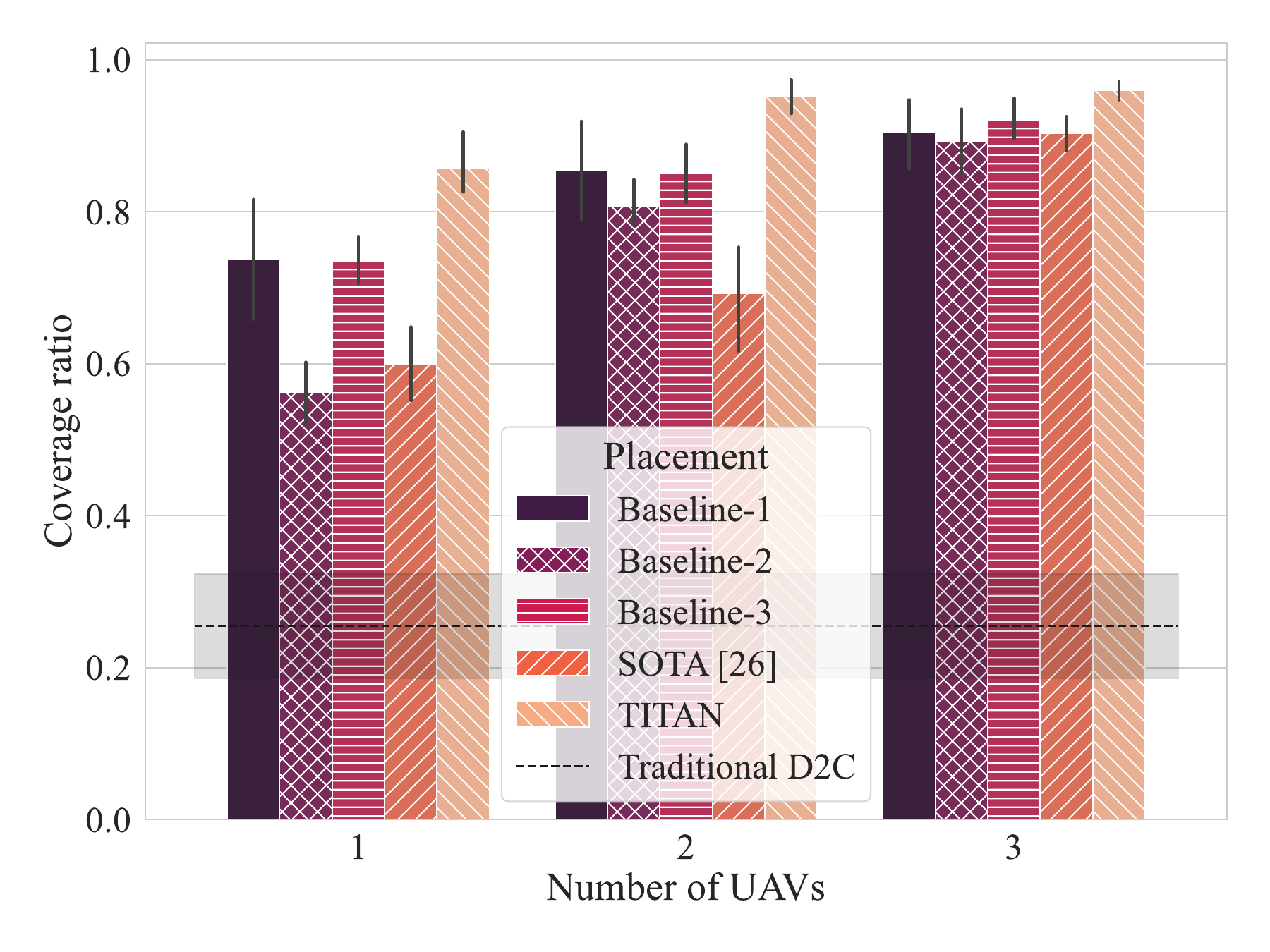}
        \caption{Coverage ratio comparison.}
        \label{fig:qos_coverage}
    \end{subfigure}
    \caption{Performance comparison of UAV placement methods in terms of sum-rate and coverage ratio. Sum-rate and coverage achieved by different UAV placement strategies as a function of the number of UAVs, demonstrating that the proposed TITAN optimization consistently outperforms state-of-the-art~\cite{multidt2025} (+32.2\% user coverage, +64.9\% sum-rate) and other baselines.}\vspace{-8pt}
    \label{fig:qos_results}
\end{figure*}

\section{Experimental Analysis} 
In this section, we present a comprehensive performance evaluation of the \name~framework based on the methodology outlined in Section V. We conduct a series of targeted experiments designed to validate \name's core contributions, analyzing its performance against established baselines and testing its robustness under non-ideal conditions. All results are averaged over 20 independent simulations and given with the 95\% confidence intervals. The codebase for TITAN framework is publicly available at~\cite{twist}.

\begin{figure}[hbt]
    \centering
    \includegraphics[width=0.45\textwidth]{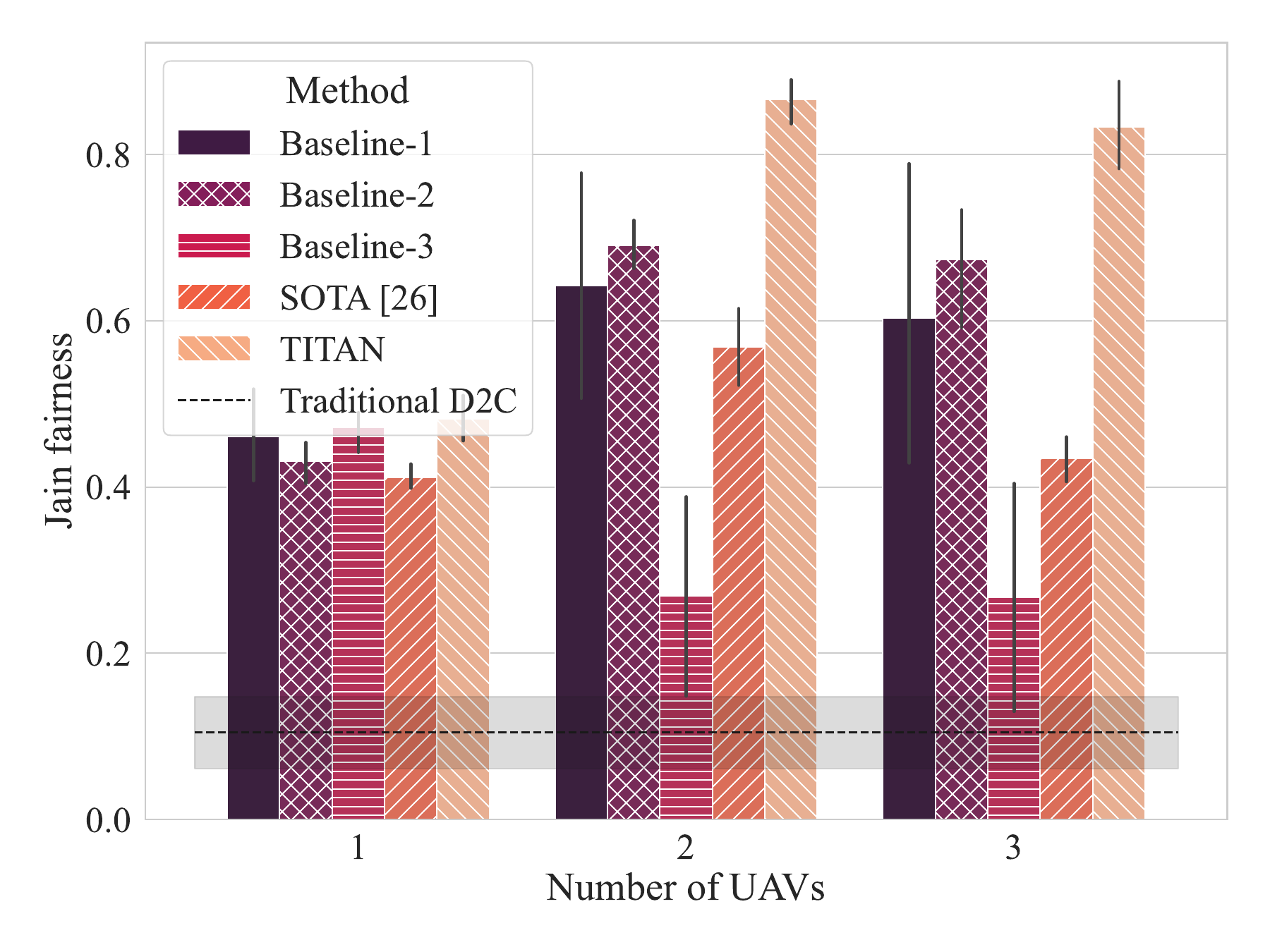}\vspace{-10pt}
    \caption{Jain's fairness index illustrating how fairness across UEs varies with increasing UAV count. Although adding UAVs initially improves fairness significantly, deploying more than two UAVs reduces fairness due to uneven distribution of excess capacity.}
    \vspace{-10pt}
    \label{fig:fairness}
\end{figure}

\subsection{Scenario 1: Baseline Performance in Full Infrastructure Collapse}

This initial experiment evaluates the fundamental effectiveness of \name~by simulating a worst-case disaster scenario where all three terrestrial gNBs have failed. This allows us to directly measure the benefits of \name's high-fidelity modeling \textbf{(C1)} and its efficient topology optimization \textbf{(C2)} in enabling D2C communication from a clean slate. Fig.~\ref{fig:qos_sumrate} and Fig.~\ref{fig:qos_coverage} illustrate the performance comparison of different UAV placement methods for LAWN with Traditional D2C.

Fairness analysis (Fig.~\ref{fig:fairness}) also demonstrates \name's superior performance. We observe a drop in fairness beyond two UAVs due to redundant coverage by additional UAVs, benefiting only a subset of users and thus decreasing overall fairness. This effect is most pronounced for Baseline-3 as maximizing a single link-quality objective tends to concentrate UAVs around already-strong links, which inflates throughput for a subset of UEs while degrading Jain’s fairness and leaving weakly served regions under-covered.

\noindent \textbf{$\bullet$ Observation 1:} \name{} consistently outperforms the SOTA and all baseline methods across all KPIs, underscoring the necessity of DT-grounded, site-specific optimization \textbf{(C1, C2)}. Overall, \name{} achieves $64.9\%$ higher sum-rate than the SOTA, improves coverage by $32.2\%$, while providing $49.3\%$ more fair service. The performance gap is more pronounced at lower UAV counts, because \name{} uses DT-informed, site-specific propagation knowledge to allocate scarce UAV resources to blockage-critical user clusters and to select 3D placements that jointly balance coverage, throughput, and fairness under the actual urban environment. In contrast, other methods rely on simplified or probabilistic channel assumptions and/or SINR-centric placement logic, which often misallocates UAVs by over-serving already-strong links and missing deterministic blockages, leading to lower effective coverage and unfair service. As the number of UAVs increases, the deployment becomes less resource-constrained and competing methods recover some QoS, but \name{} maintains a clear advantage by avoiding redundant coverage and consistently optimizing the coverage--capacity--fairness trade-off via DT-guided placement.

\begin{figure*}[hbt]
    \centering
    \begin{subfigure}[b]{0.45\textwidth}
        \centering
        \includegraphics[width=\textwidth]{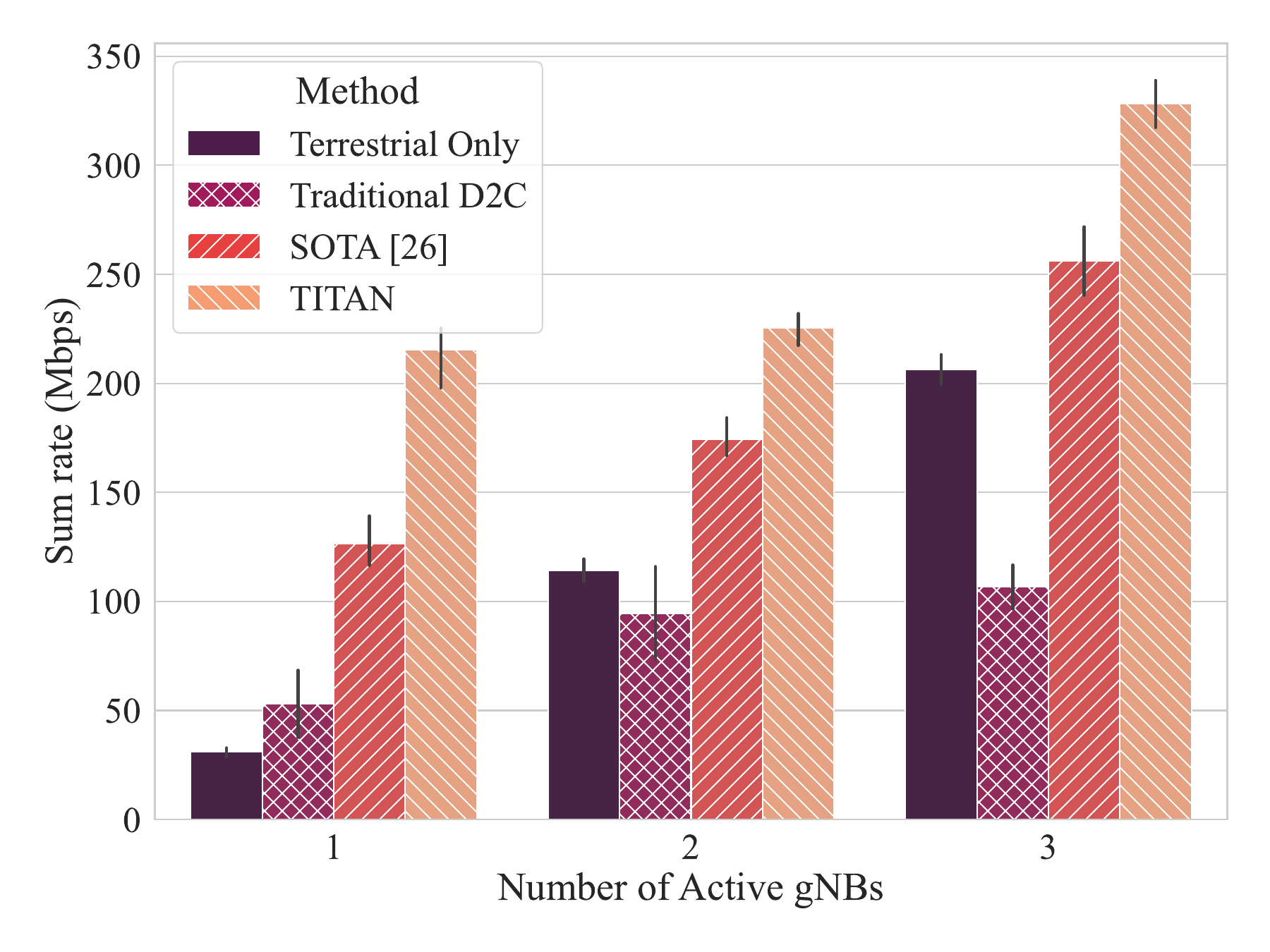}
        \caption{Sum-rate comparison.}
        \label{fig:fail_sumrate}
    \end{subfigure}
    \begin{subfigure}[b]{0.45\textwidth}
        \centering
        \includegraphics[width=\textwidth]{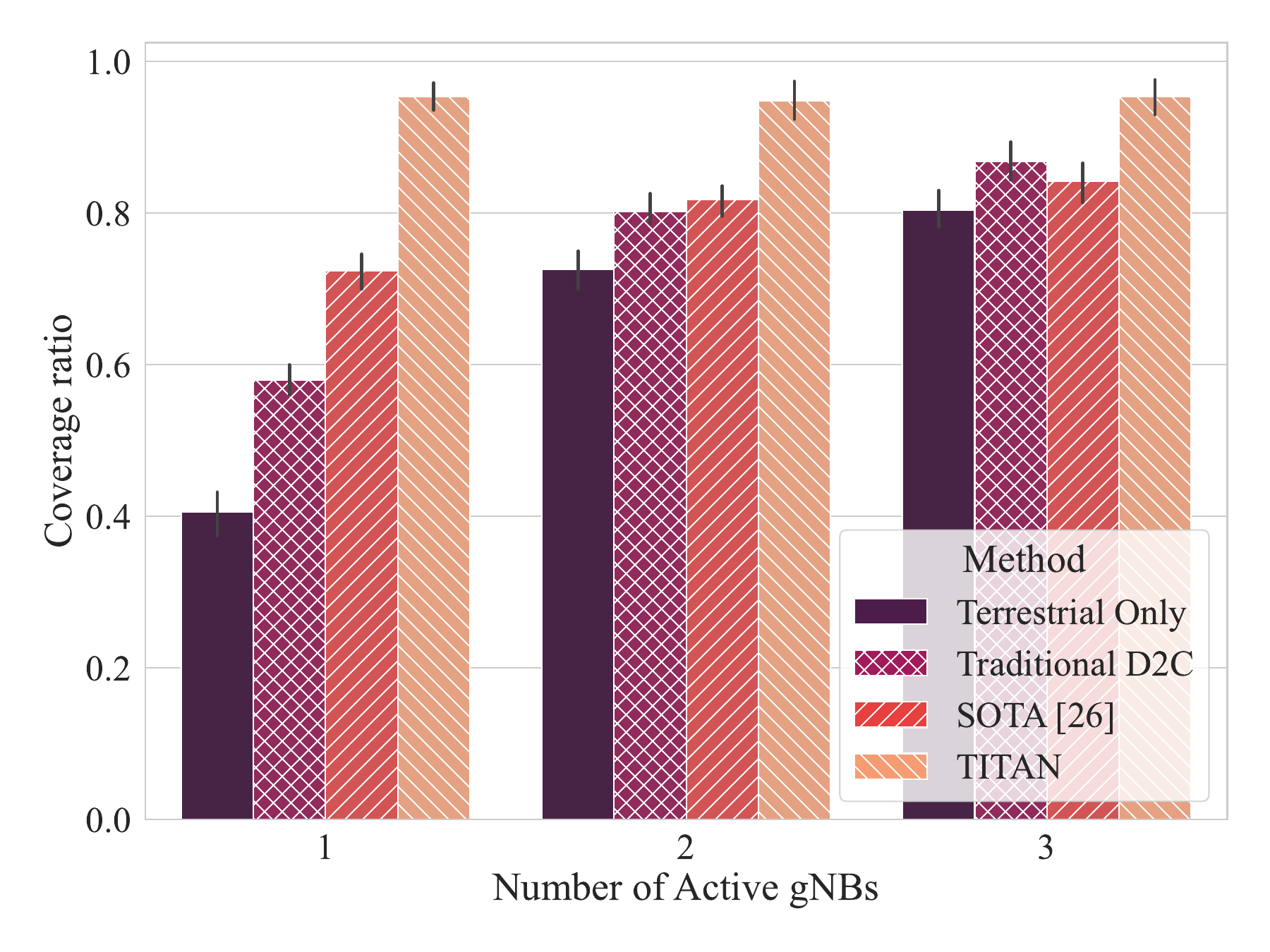}
        \caption{Coverage ratio comparison.}
        \label{fig:fail_coverage}
    \end{subfigure}
    \caption{Disaster adaptability analysis. Sum-rate and coverage achieved by scenarios considering active gNB count. This demonstrates that the proposed \name~provides the best disaster response while considering coverage.}
    \vspace{-10pt}
    \label{fig:fail_results}
\end{figure*}

\subsection{Scenario 2: Dynamic Adaptation to Partial Failures}

This experiment is designed to validate \name's ability to reactively adapt its topology to a dynamic disaster environment \textbf{(C3)}. The simulation begins with all three terrestrial gNBs active, providing baseline coverage. We then simulate cascading failures by disabling the gNBs one by one. After each failure, \name~is triggered to re-optimize and deploy an aerial topology to restore service. To enable a fair disaster-response comparison against~\cite{multidt2025}, we generate the ground-user population around the three gNB sites and treat these sites as the mission-relevant AoIs.

We compare deployments with various BS failure scenarios (0, 1, and 2 BS outages, meaning 3, 2 and 1 gNBs are active) against scenarios with Traditional D2C and \name~assistance to demonstrate effectiveness clearly in Fig.~\ref{fig:fail_sumrate} and Fig~\ref{fig:fail_coverage}. Overall, \name~consistently outperforms the other methods across all outage cases by jointly accounting for the site-specific, DT-informed re-optimization.

\noindent\textbf{$\bullet$ Observation 2:} \name~demonstrates highly effective reactive adaptation. After the initial gNB failure, \name's deployment of a single UAV restores over 80\% of the original network sum-rate. With subsequent failures, \name~continues to compensate by deploying additional UAVs, successfully maintaining a high level of service continuity. Although Traditional D2C restores the initial coverage, it cannot fully restore the required sum-rate. This validates the framework's ability to dynamically adapt to evolving on-the-ground conditions \textbf{(C3)}.

\subsection{Scenario 3: Robustness to Imperfect Digital Twin}

This final set of experiments evaluates \name's resilience to the inevitable imperfections of a real-world deployment, directly validating its robustness \textbf{(C4)}.

\begin{figure*}[hbt]
    \centering
    \begin{subfigure}[b]{0.45\textwidth}
        \centering
        \includegraphics[width=\textwidth]{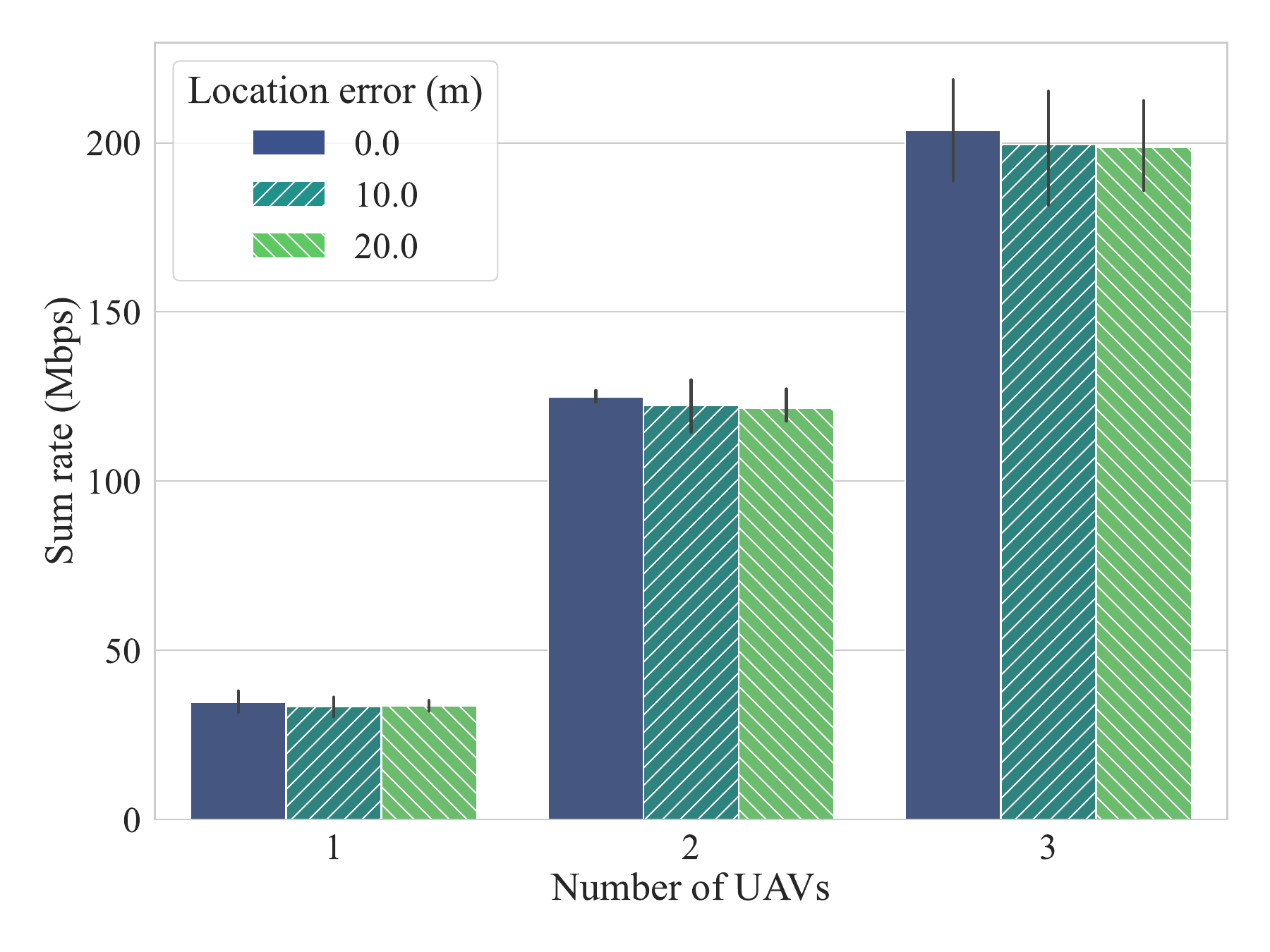}
        \caption{Sum-rate vs. DT accuracy}
        \label{fig:acc_sumrate}
    \end{subfigure}
    \begin{subfigure}[b]{0.45\textwidth}
        \centering
        \includegraphics[width=\textwidth]{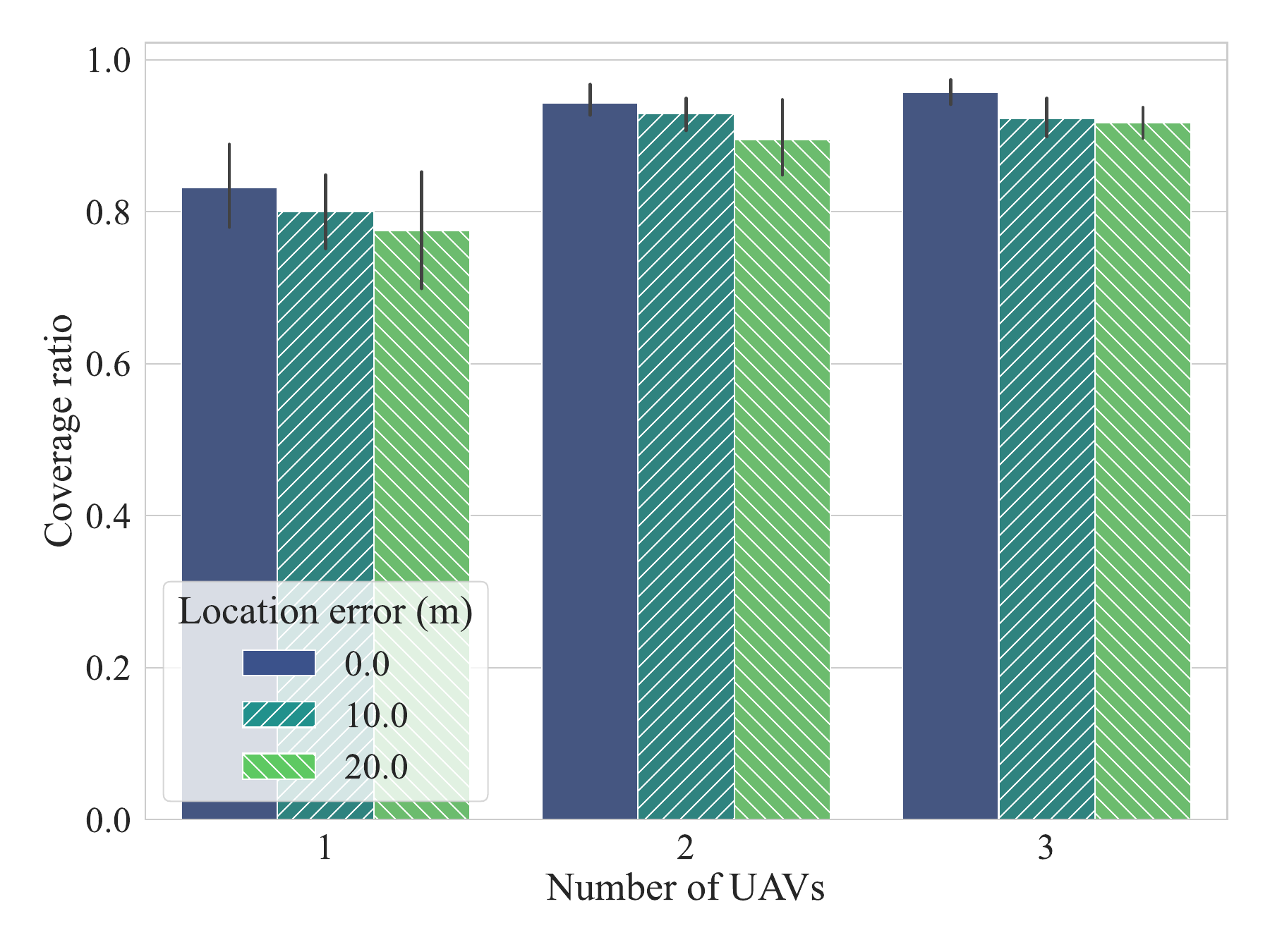}
        \caption{Coverage ratio vs. DT accuracy.}
        \label{fig:acc_coverage}
    \end{subfigure}
    \caption{Impact of UE localization accuracy on UAV placement optimization outcomes. Demonstrating QoS degradation as UE location estimation errors increase.}\vspace{-10pt}
    \label{fig:accuracy_results}
\end{figure*}

\subsubsection{Impact of UE Location Error}

Initial QoS evaluations assumed perfect UE location knowledge, an unrealistic assumption in practical scenarios. To address this, we evaluate how localization errors influence placement performance. Results confirm that as UE location error increases it also slightly reduces the effectiveness of \name~due to the mismatch between the estimated and actual channel responses. This results in suboptimal UAV placement for the air layer.

\begin{figure}[hbt]
    \centering
    \begin{subfigure}[b]{0.49\linewidth}
        \centering
        \includegraphics[width=\linewidth]{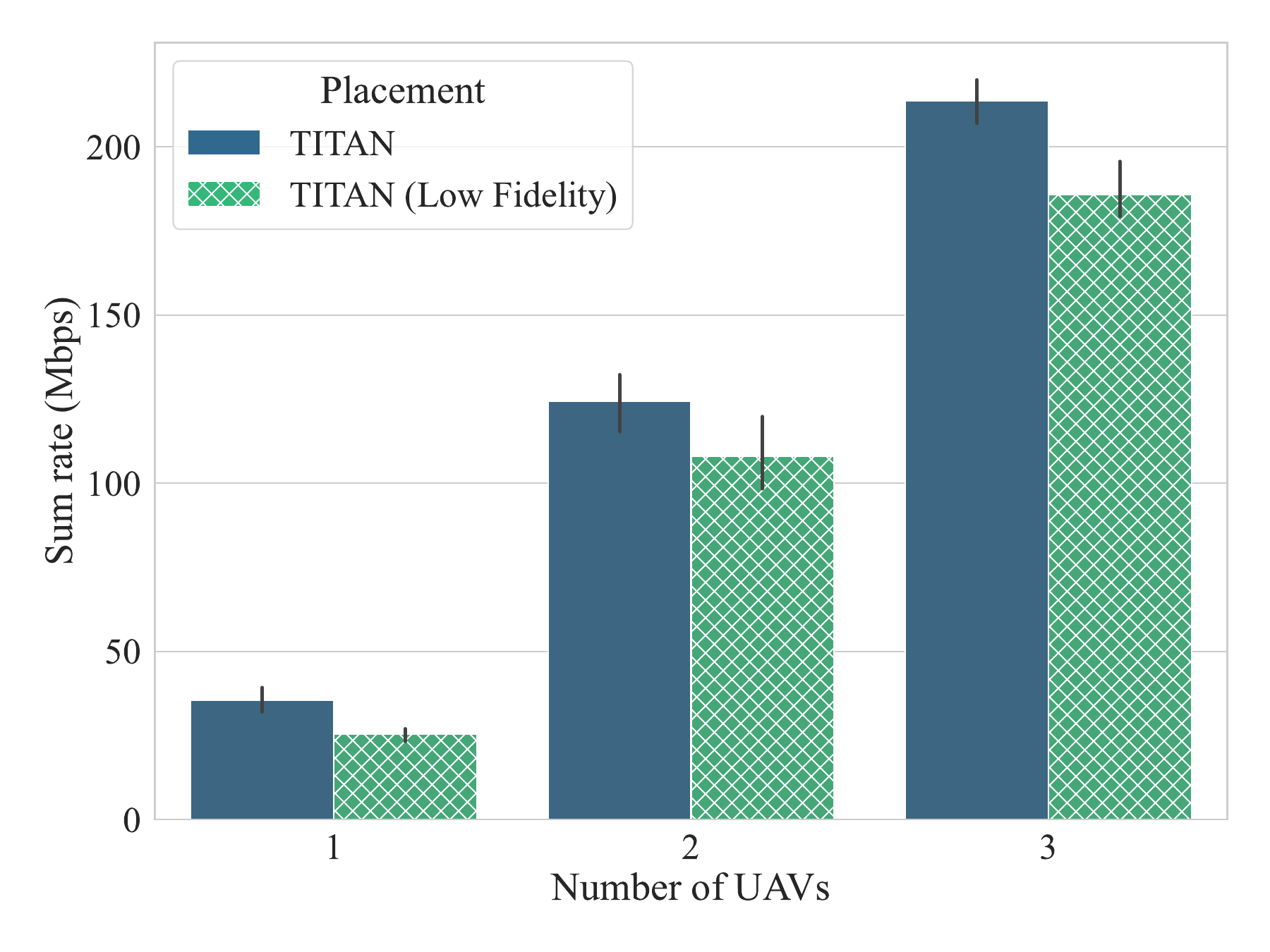}
        \caption{Sum-rate vs. fidelity level.}
        \label{fig:sum_fidelity}
    \end{subfigure}
    \begin{subfigure}[b]{0.49\linewidth}
        \centering
        \includegraphics[width=\linewidth]{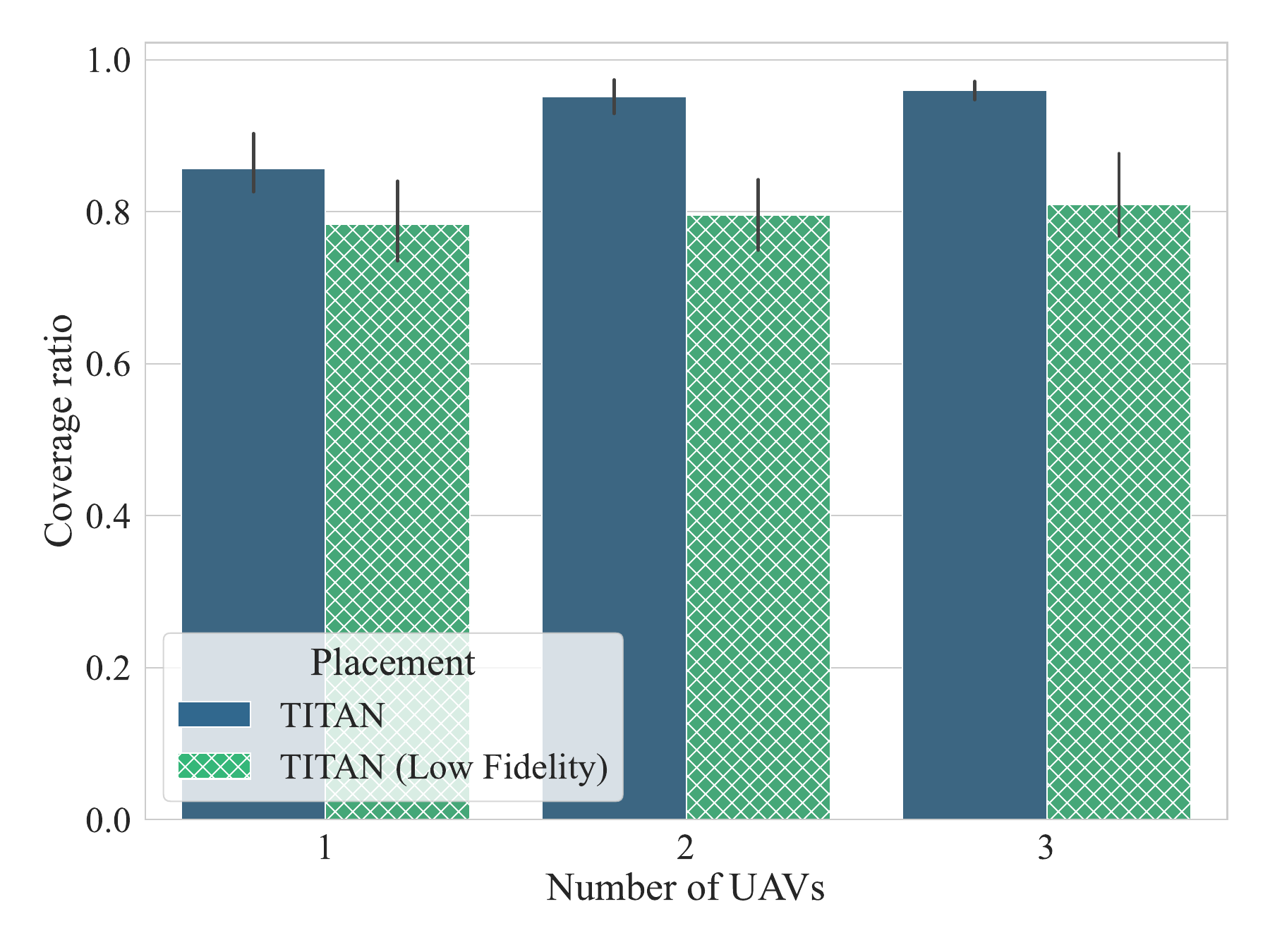}
        \caption{Coverage ratio vs. fidelity level.}
        \label{fig:cov_fidelity}
    \end{subfigure}
    \vspace{-5pt}
    \caption{Effect of ray-tracing fidelity on QoS metrics. Low-fidelity ray-tracing severely underestimates inter-UT interference, reducing overall system throughput and coverage compared to high-fidelity simulations.}
    \vspace{-10pt}
    \label{fig:fidelity_results}
\end{figure}

\subsubsection{Impact of DT Fidelity}

We evaluate how varying the fidelity level of ray-tracing simulations impacts UAV placement optimization. Specifically, we compare the standard high-fidelity mode of \name~against a low-fidelity configuration that creates only \textbf{1\% of the rays} per transmitter. While the low-fidelity mode reduces computation time by nearly half per iteration, it significantly underestimates interference and coverage, leading to notably inferior placement outcomes compared to the full-fidelity version.


\noindent \textbf{$\bullet$ Observation 3:} 
Optimizing with a lower-fidelity DT reduces the time to complete all $N_{iter}=150$ optimization iterations from $\sim\!23.1$\,s to $\sim\!12.9$\,s (approximately $1.8\times$ faster) on an RTX 4090 GPU, while still achieving over 85\% of the full-fidelity sum-rate.
Crucially, in both low- and high-fidelity modes, TITAN converges to near-optimal placements with fewer optimization steps than the state-of-the-art, because the DT provides a better-informed and environment-consistent objective landscape. Even in low-fidelity mode, the search remains physics-grounded rather than proxy-grounded (e.g., stochastic LoS/SINR surrogates), preserving most of TITAN’s advantage while substantially reducing computational cost. The DT construction and optimization run centrally (e.g., on an incident-command edge server), and only the resulting UAV placement/topology commands are pushed to the UAVs over the control link.
This demonstrates that \name{} can provide a strong, near-optimal solution even under tight computational constraints, further supporting practical robustness \textbf{(C4)}.

\subsection{Data Collection Capabilities}

The \name~framework is equipped with comprehensive data collection capabilities that extend well beyond conventional metrics. It enables the extraction of fine-grained coverage maps, full channel impulse and frequency responses. This depth of data is instrumental for meticulous deployment planning, system performance analysis, and robust contingency preparation in disaster-prone scenarios. Beyond its immediate operational benefits, \name~serves as a powerful enabler for data-driven wireless communication research. By enabling to collect channel impulse responses from multi-antenna UAVs at diverse spatial configurations, the framework provides an invaluable resource for the training and validation of advanced methods such as CSI-based localization algorithms and machine learning models. These extensive datasets foster the rapid prototyping and testing of next-generation wireless solutions, thereby accelerating innovation and agility within the research community~\textbf{(C5)}.

\section{Conclusions}

In this paper, we introduce \name, a novel framework that enables resilient D2C communication in urban disaster zones by leveraging a LAWN architecture with an adaptive aerial topology. We demonstrated that by integrating ray-traced channel evaluation (Sionna RT) into a DT-based optimization loop, \name~drastically outperforms approaches relying on inaccurate stochastic channel models, achieving deterministic, site-specific propagation predictions that are critical for reliable UAV placement in complex urban environments. This physics-based channel knowledge enables Bayesian optimization to efficiently find near-optimal UAV placements that intelligently adapt the network topology to overcome the severe blockages rendering traditional D2C nonviable. Our extensive system-level simulations show that \name~dramatically outperforms both baseline approaches and the state-of-the-art, as it significantly increases network coverage and capacity, while providing fairer service. We validated \name's robustness against real-world imperfections, such as UE location errors and varying DT fidelity. For future work, we plan to enhance the DT's fidelity by incorporating environmental factors like weather models and exploring the possibility of creating candidate rays according to antenna radiation patterns to fully ray-trace the D2C backhaul link. We release TITAN framework as an open-source tool to enable reproducible research and foster community-driven advancements in creating the resilient communication networks of the future.

\section*{Acknowledgement} 
This work is supported by Istanbul Technical University, Department of Scientific Research Projects (ITU-BAP, 45375) and the authors gratefully acknowledge the support from NVIDIA Academic Grant Program.


\bibliographystyle{IEEEtran} 
\bibliography{reference}

\end{document}